# On the quality of the Olmec mirrors and its utilization

José J. Lunazzi

Universidade Estadual de Campinas - Instituto de Física

13084-970 - Campinas - SP - Brasil

lunazzi@ifi.unicamp.br

## ABSTRACT

Archaeological mirrors from the Olmec civilization were analyzed in the context of bibliography produced in the last two decades. Photographs of its images are showed as a proof of its good quality. Some suggestions are made on its probable utilization.

## 1. INTRODUCTION

It is the purpose of this article to synthesize the ideas sometimes maybe naive, sometimes very technical, that an expert in optics may have when consulting the references more easily available on the more ancient archaeological mirrors from Mexico. Olmec mirrors were dated by the radiocarbon method by Heizer[1] as old as 3125 years up to 2130 years from now. They were used, as some archaeologists suggest, for making fire, self contemplation, medicine, divination and astronomy. The range of focal lengths goes from 5cm to more than 80cm, while a few convex mirrors are known. Many examples of their appearance in iconography are related to the god sun or to its utilization as a pectoral or in the back probably being a symbol of high social status. The older mirrors dated in America are from the Incas, about 800 years before the Olmecs, and their appearance goes at least to the Teotihuacan civilization, a few centuries before the colonization. The extension of the geographic area where they existed seems to us not entirely well known, since those made on pyrite, maybe the more reflective material, deteriorates very easily making difficult to be recognized as mirrors. Besides that, for civilizations living at the jungle, the extension and difficulties of the territory makes difficult to obtain archaeological material. Being associated to the polishing of stones with reasonable sphericity, would be interesting to say that the appearance of some ovoidal stone artifacts was reported at a site in Mogi Mirim city, Sao Paulo State, Brazil.

## 2. IMAGES OBTAINED WITH THE MIRRORS

The quality of Olmec mirrors is very high as it can be seen from the images they generate. To our knowledge, the only images generated with them that were published comes from Heizer[1] in Fig.6, with an object at short distance from the mirror. Saunders[2] showed in Fig.1 the image of an object reflected at an Inca mirror whose curvature can be estimated as plane or almost plane because no magnification can be observed on the image. We show in Fig.1 of this paper a similar result but obtained from a convex Olmec mirror whose diameter is about 5cm with object and camera in front of it at a 35cm distance.

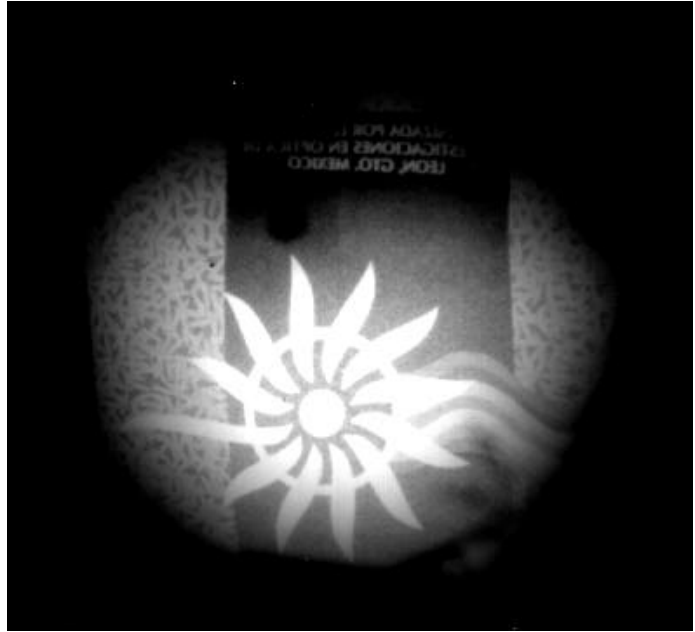

**Figure 1:** Image of printed letters and color symbol reflected from an Olmec Mirror. Photograph by the author.

Since the quality of mirror images depends on the distance from object to mirror, we show in Fig.2 the acchieving of an image coming from solid letters located at about 9m distance, in a situation where the light impinged at a 45° angle, approximately. The negative was inverted from right to left  when enlarging to the paper copy, that is way the letters are not inverted and can be read.

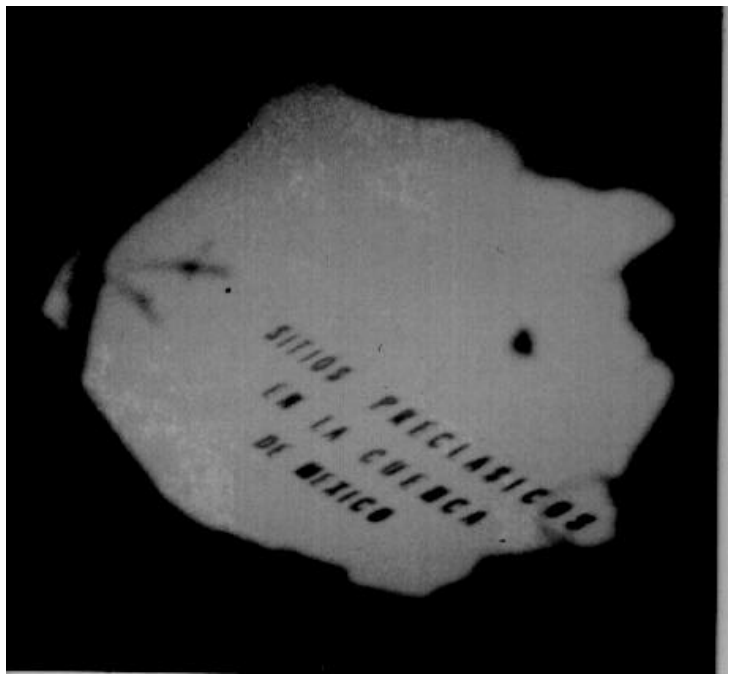

**Figure 2:** Reflected image of moulded metal letters located at 9m distance. Photograph by the author.

These are good images although we could not work at ideal conditions: the focal length of a 105mm objective was doubled by means of a diverging additional lens, and we were not allowed to remove the protective glass, which was traversed twice by the light. KODAK black and white negative film TMAX 400 was employed and developed in an improvised darkroom. A film of more resolution and better laboratory conditions would be desirable. We photographed both images from the only mirror exhibited at the Pre-Classic Olmec section of the "Museo Nacional de Antropologia" in Mexico city.

Since convex mirrors are very rare, it is useful to say that this mirror is not the one reported by Carlson[3] in p.123, what we determine because of its different shape. It is convex having about 50cm focal length, what we calculated by measuring the focusing conditions at the image and with classical laws of image formation. Convexity on the mirrors could be a practical way of obtaining the whole image of a face in a element of reduced size for using it, i.e., in cosmetic. With some concave Olmec mirrors we could photograph good quality real images, being clear that a hand or even a face can be binocularly seen as floating in the air at about 8-16cm distance in front of the mirror and inverted. They are very impressive and we can think on how strong feelings the images could have been caused to the Olmec people. Our knowledge of the Olmec iconography is very little, but we can think that upside inverted images should be present on it. We want to mention Fig. 35 in a paper by Carlson[3] where the Tablet of the Foliated Cross was drawn and one symbol resembling a human face can be seen inverted below the main figure. It appears three times at the center, as a smiling character. Two larger ones can be seen at both sides of them. In Fig. 37 of the same article a similar (but not smiling) symbol appears twice below the main figure.

The very well known sculptures of the feathered serpents at Teotihuacan are described as exiting from mirrors by Taube[4], so we also mention here its similarity to a well known experience of locating the image of a flower just above a jar to give the illusion of a realistic but untouchable flower, which was known as "La Rosa Azteca" (The Aztec Rose) in some popular attractions in Argentina a few decades ago[5] .

Plane mirrors are not reported on the articles we know but they were probably employed[1]. Would axially symmetric figures be an evidence of that? There is a painting in the front wall of a construction at Teotihuacan site with the figure of two jaguars, many meters long, perfectly symmetric one another, tail by tail. The utility of plane mirror could also be for making light signals[6], since plane mirrors do not add divergence to the sun light.

The possible use of concave mirrors for generating a projected image ("camera obcura") could not be tested by us, but it is unprobable that mirrors with two focal lengths which may be toroidal could do that task because of its strong astigmatism. Only spherical or double cylindrical mirrors of the same focal length can be used for projecting images on a screen or making fire. The astigmatism is a result of having two different focal lengths[1,2] and appears very reduced in visual observations because the pupil size of the eye is small enough to impose restrictions on it.

According to Carlson[3] the sphericity of the mirrors can result as a natural consequence of the grinding process, but some more refined technique based on the observation of the image quality would not be discarded. Mosaic mirrors up to 50cm in diameter are referred by Marshac[3] and we think that for assembling them an optical technique would be much more effective than any attempt of purely mechanical control. The only utility of making a greater curvature in one direction would be to enlarge the field of viewing. This seems to happen always in the vertical direction, considering the mirror being used as a pectoral with two lateral holes for fixation. In that case, the mirrors could had being used to show the image of the sun to a larger audience, while tilting the breast for horizontally scanning them. The image of a part of a circular object reflected at different tilted positions is clear to us when analyzing different representations of mirrors reported by Carlson [3] (i.e. Fig. 34).

## 3. THE USE OF MIRRORS FOR FIRING PURPOSES

In respect to their use for fire making purposes, the only successful experience was obtained with one spherical mirror by Ekholm[3] in 1973, probably using dry wood, reported in a paper at the "Congreso degli Americaniste" in Rome. That article could not be found neither by us nor by other archaeologists[1], who also tried to obtain fire, unsuccessfully. A description of the operation was not available to us. The Inca Garcilaso de la Vega wrote, in his "Libro Sexto de los Comentários Reales de los Incas", ch. XXII, that the way the Incas made sacred fire was by using a "highly burnished concave bowl, in the shape of a half orange, and where the sun rays concentrated, they put a piece of not burned cotton, which is very flammable". Garcilaso wrote in old Spanish, an intricate language of Cervantes's times. Concave bowls similar to the ones reported by Garcilaso can be found at a collection in Lima, Peru, made on gold[7]. Nordeskiold[3] and also Cooper[3] expressed serious doubts about Garcilaso's report. In our consultation to bibliographic references we did not found any mention to the "not burned cotton" technique, which seems to be of great help in making fire because it dries and blackens the cotton, making it very absorbing to the luminous energy.

The reflectivity of the actual mirrors is not reported in the references but that of polished materials is 21% for Magnetite, 28% for Hematite[8] and 55% for Pyrite. Although we do not know if the Olmecs would had employed the "not burned cotton" technique of the Incas, we made successful simulated experiences in this way using a spherical mirror of 50mm focal length, 15mm diameter. aluminized to 80% reflectivity. We made it twice, one in Campinas city, Brazil, which is at the latitude of the tropic of Capricorn, and the other in Mexico city, both in September (not summer) days about 24º C temperature with some wind. Our results could be compared to the case of Olmec mirrors using a 20% reflectivity spherical mirror of 30mm diameter, or a little larger if we consider the expansion of the image of the sun by spherical aberration. It was clear that only the blackened cotton can start fire, not possible with not prepared cotton, and that by using dry leaves or wood we could only obtain smoke. Taube[4] shows the identification of mirrors with cotton, and also the identification of darts with cotton. Would the darts be used with cotton for launching fire? That is, would cotton be employed as a firing material?

## 4. ON THE QUALITY OF REFRACTIVE ELEMENTS

Very well polished quartz objects are known to belong to the Olmec culture. Among them there is a small sphere, smaller than 10mm in diameter, within the group of offerings of the E tomb at La Venta, Tabasco, in the permanent exhibit of the "Museo Nacional de Antropologia" at Mexico city. Although we could not see an object through it, its appearance made us to think that this element may function as a magnifying lens.

## 5. THE APPLICATION OF HOLOGRAPHY IN ARCHAEOLOGY

Holography is a well known technique invented in 1947 as a "wavefront reconstruction technique" by Dennis Gabor, who received the Nobel prize for it. It was already employed for registering complete detailed images of archaeological objects, to record the structure of a mummia at the very moment of opening it and, in an another case, to allow the general public to see the image of the skull of an ancestor of more than one million years old[9].

Holographic interferometry is being employed for evaluation of ancient material and for restoration purposes[10].

We propose its use for registering in a single exposure the complete optical characteristics of an archaeological object to be analyzed at the optical laboratory reaching an interferometric level of analysis.

Its advantages can be:

i) non-destructive testing, since the metallic sphere of a dial gauge needs not go over its surface. The necessary laser exposure is very low, much less than its exposure to the usual light of the ambiance.

ii) more complete geometrical description of the surface, because any direction can be tested and contour lines determined.

iii)  more information on the structure of the surface because interferometry can be applied to sub-micron resolution identifying roughness to a level that microscopy can not reach.

An adaptation of a didactic equipment[11] seems to be a simple way of making small holograms when the Museum does not have enough space for mounting a holographic laboratory, neither a darkroom. A portable version of it is usually carried by us in a simple rigid suitcase.

## 6. CONCLUSIONS

From the information we obtained it is clear that Olmec mirrors are very valuable cultural elements of a quality not known to the optical community neither to the general public. The precise characterization of its optical parameters seems not being completely reported, so that its measurement can be an interesting subject of  research. The shape of the optical surfaces and its imaging performance may deserve further analysis. We propose the use of holography to record a wavefront that can be reproduced at the optical laboratory for measuring the shape of the mirrors and also performing interferometry to measure the quality of the surfaces. Seems to us that the joint work of archaeologists and members of the optics community would contribute to a better knowledge of the subject.

## 7. ACKNOWLEDGEMENTS

The author wish to express his acknowledgment to the Organizing Committee of the "II Reunión Iberoamericana de Óptica" , Guanajuato - GTO, Mexico, 18-22 September 1995 for their invitation to present this work at the meeting, particularly to its General President Dr. Daniel Malacara Hernandez whose wide way of thinking and acting has allowed the realization of many important activities in Latin America.

Financial support from FAEP - Campinas State University, FAPESP - Foundation for Assistance to Research of the Sao Paulo State, CNPq - National Council of Research made possible this work. I am also grateful to Jesus Najera who, representing the Esperanto world community, helped with equipment and personal work to obtain the photographs. The "Museo Nacional de Antropologia" of Mexico city is acknowledged for allowing  to introduce the photographic and video cameras, its mechanical holders and a lamp in their exhibiting hall to perform the photographs. Alexsandra Siqueira is acknowledged for helping at the photographic laboratory. A. Pimenta Lima is acknowledged for donating one archaeological embossed hologram[9] for evaluation.

## Comments after edition:

This article is a pre-print version of the one published in the Proceedings of the "II Reunión Iberoamericana de Óptica", Guanajuato - GTO - Mexico, 18-22 Sept. 1995, SPIE V 2730, p.2-7.

After reviewing the article, the author found that the interpretation of the text of Garcilaso de la Vega as "not burned cotton technique" may not be precise, and if so, the technique of employing a piece of fire-blackened cotton for firing purposes must be interpreted as a possibility suggested by Lunazzi itself.

See also:

"*Olmec Mirrors: an Example of archaeological American mirrors*", J.J. Lunazzi, chapter of the book "Trends in Optics" V3 to be published by the International Commisssion for Optics - ICO, Ed. Anna Consortini, Ac. Press. 1996, p411-421, including a color picture.